\documentstyle[12pt]{article}
\def\frac#1#2{ {{#1} \over {#2} }}
\def\ie{\hbox{\it i.e.}{ }}

\def\half{\mbox{\small $\frac{1}{2}$}}

\def\re#1{(\ref{#1})}
\def\beq{\begin{equation}}
\def\eeq{\end{equation}}
\def\beeq{\begin{eqnarray}}
\def\beeqn{\begin{eqnarray*}}
\def\eeeq{\end{eqnarray}}
\def\eeeqn{\end{eqnarray*}}
\def\se{S_{\mbox{\footnotesize{eff}}}}
\def\ser{S_{\mbox{\footnotesize{eff,rel}}}}
\def\Pirel{\Pi_{\mbox{\footnotesize{rel}}}}
\def\set{S_{\mbox{\scriptsize{eff}}}}
\def\si{S_{\mbox{\footnotesize{int}}}}

\def\Girr{\G_{\mbox{\footnotesize{irr}}}}
\def\De{\D_{\mbox{\footnotesize{eff}}}}
\def\Der{\D_{\mbox{\footnotesize{eff,rel}}}}

\def\DG{\D_{\G}}

\def\bchi{\bar \chi}

\def\bpsi{\bar \psi}

\def\s{\sigma}

\def\G{\Gamma}
\def\eps{\epsilon}
\def\L{\Lambda}

\def\D{\Delta}
\def\d{\delta}

\def\LdL{\L\partial_\L}
\def\UV{$\L_0\to\infty\;$}

\def\bit{\begin{itemize}}
\def\eit{\end{itemize}}
\def\ben{\begin{enumerate}}
\def\een{\end{enumerate}}

\def\nome#1{{\label{#1}}}
\def\p{\partial}
\def\K{K_{\L\L_0}}

\begin{document}
\begin{titlepage}
\begin{flushright}
SHEP 96-08 \\
UPRF-96-458 \\
hep-th/9602156
\end{flushright}
\vspace{.4in}
\begin{center}
{\large{\bf Gauge Invariance, the Quantum Action Principle, and the
Renormalization Group}}
\bigskip \\ Marco D'Attanasio$^{1,2}$ 
and Tim R.  Morris$^1$ 
\\
\vspace{\baselineskip}
{\small 1. Physics Department, University of Southampton,
Southampton SO17 1BJ, UK}\\ 
{\small 2. I.N.F.N., Gruppo collegato di Parma, viale delle Scienze, 
43100 Parma, Italy} \\
\mbox{} \\
\vspace{.5in}
{\bf Abstract} \bigskip \end{center} \setcounter{page}{0}
If the Wilsonian renormalization group (RG) is formulated with a cutoff that
breaks gauge invariance, then gauge invariance may be recovered only once
the cutoff is removed and only once a set of effective Ward identities
is imposed.  
We show that an effective 
Quantum Action Principle can be formulated in perturbation theory 
which enables the effective Ward identities to be solved order by order, 
even if the theory requires non-vanishing subtraction points. 
The difficulties encountered with non-perturbative approximations are 
briefly discussed. 
\end{titlepage}

\section{Introduction}
In trying to formulate a non-perturbative 
RG \footnote{\ie the Wilsonian RG, also called the exact RG.}
for a gauge theory,
we encounter the problem that the division of momenta into large or small
(according to some scale $\Lambda$) -- which is fundamental to the exact RG
approach -- is incompatible with gauge invariance. This is easy to
appreciate if one considers a homogeneous gauge transformation $\Omega$
on some 
matter fields $\Phi(x)$,
\beq\nome{aa1}
\Phi(x)\mapsto \Omega(x) \Phi(x) \,.
\eeq
Since in momentum space, $\Phi(p)$ is mapped into a convolution with the
gauge transformation, any division of momenta into high and low is seen
not to be preserved by gauge transformations. 
In order to solve this problem, clearly we are presented with two options:

(a) We break the gauge invariance in intermediate steps, and aim to recover 
the gauge 
invariance once $\Lambda$ is removed -- by imposition of some constraints.

(b) We generalize the RG in such a way that it manifestly preserves the
gauge invariance.

Clearly, manifest preservation of gauge invariance would be preferable. 
Unfortunately for all but the very simplest case of pure U(1) gauge
theory \cite{u1}, 
we encounter in option (b) the well known problem that it is not easy
to regularise non-perturbatively while preserving both Poincar\'e invariance
and gauge invariance. Actually the problem is worse than that \cite{u1}, 
because the resulting RG must also allow for manageable approximations,
and this has so far required that the cutoff be effectively placed inside
a free propagator, breaking the gauge invariance for all but pure U(1)
gauge theory. We will not discuss this option further here, but instead 
give a detailed study of option (a). 

Option (a) can be carried out to all orders in perturbation theory,
relatively straightforwardly, since, as we will show, an analogue of the
Quantum Action Principle (QAP) exists for the solution of the broken
Ward identities. When these are solved, the unbroken Ward identities are
guaranteed to hold once the cutoff is removed.
Indeed in this way, renormalized physical Green functions, 
with the correct gauge dependence, evaluated
at non-zero subtraction points (when necessary, \ie when massless particles
are present), may be
constructed  order by order in the couplings. Our study of the QAP also
serves as a basis to study non-perturbative approximations. We point out
the difficulties involved in finding viable non-perturbative 
approximations, \ie ones in which the appropriate
Ward identities are exactly obeyed.

Some comments on the use of the background field method are now necessary.
If the RG is combined with the background field method \cite{RW,newRW}, then 
background gauge invariance may be maintained by replacing the division
of momenta into high or low, by a division of eigenvalues of the 
background covariant Laplacian. The problems of option (b) above are
not encountered here because the background field does not propagate.
Equally however, note that nor
does background field invariance replace the need to ensure 
that the quantum gauge invariance\footnote{\ie BRS invariance in gauge
fixed systems} is respected by the quantum fields, for example 
it is the latter that ensures that longitudinal modes are properly cancelled
by ghosts in internal propagators, \ie that unitarity is maintained, not
the former. Thus these methods must also be treated according to
category (a).

Lastly, let us  
stress that it is crucial that 
approximations of the RG do not {\sl of themselves} result in breaking
of the (quantum) gauge invariance in the final answer.
This requires, for (a), exact preservation of the broken Ward identities.
Of course, otherwise, spurious violations 
of unitarity will be encountered, but 
many other properties of gauge theories
would also be lost. Amongst the most
important, we mention Elitzurs theorem, which in turn implies the
non-existence of local order parameters \cite{ID}, Instantons (more
generally any topologically non-trivial principal fibre bundles),
the existence of
Gribov fundamental domains,\footnote{yielding non-perturbative corrections
to the standard Ward identities\cite{Grib}} 
and anomalies. It is hard to see how these
properties would be properly incorporated if gauge invariance is only
``approximately'' conserved. Indeed it may even be the case that some of
these properties cannot be recovered at all by method (a).
In this case, non-perturbatively, Wilson RGs
of the form (a) would fail to access certain continuum limits.
(Equivalently,
all regularizations with gauge non-invariant physical cutoffs 
would fail to be in the basin of attraction of the relevant fixed points.)

Let us assume that all the modes of the fields above $\L$ 
are integrated out to generate a 
Wilsonian effective action $S(\L)$.
The dependence on $\L$ of $S(\L)$ can be translated into a differential 
evolution equation, which in general has the form
\beq\nome{1}
\p_\L S(\L)=F[S(\L);\L]\,.
\eeq
This flow equation can be used to define the theory. However, it is 
crucial to understand the symmetry properties of the effective action, \ie
of the required 
solution to \re{1}. At the effective level the local gauge symmetry of the 
theory is expressed by a set of {\it{effective Ward identities}} (see 
sect.~4)
\beq\nome{2}
\D[S(\L);\L]=0\,.
\eeq
The functional $\D$ satisfies a linear evolution equation
\beq\nome{3}
\p_\L \D= L\cdot\D\,.
\eeq
We address the problem of the locality of $\D$, \ie we 
discuss the so-called Quantum Action Principle (QAP)
\cite{QAP1,QAP2,QAP3,QAP4}, which is a major step 
towards \re{2}. The QAP will be proven in perturbation theory for a general
gauge theory, no matter if the theory contains massless particles which 
require subtraction points to define the local approximants of $\D$ and 
$S$. We will find, after Legendre transforming, that
$$
L=\hbar L_0 +{\mathcal O}(\hbar^2)\,.
$$
{F}rom this, it follows that $\p_\L \D^{(\ell)}=0$, where $\ell$ is the first 
loop order in which $\D$ is non-vanishing. Then it will be enough to show 
that $\D^{(\ell)}$ is local at a particular value of $\L$ to get the QAP
in perturbation theory.

The perturbative QAP is a well-known property in field theory. However, it is
interesting to discuss it with the modern point of view of Wilson RG method.
Moreover there have been attempts to give eq.~\re{1} a meaning which 
goes beyond perturbation theory (for instance by performing truncations of 
the effective action). 
Eq.~\re{3} can be used to study how the evolution of the couplings 
``deviates'' from the gauge invariant trajectory given by \re{1} and \re{2}.

In the next section we recall the QAP. In sect. 3 the RG flow equations
are constructed and discussed, and finally in sect. 4, the effective
Ward identities are derived and we prove that they may be cast as local
in each order of perturbation theory. The final section
concludes with some remarks on non-perturbative approximations, in particular
on the use of a derivative expansion, 
and/or truncations \cite{u1,RW,newRW,derexp,others,E}.
Apart from these final remarks, we will work in a perturbative language,
however the equations \re{1}--\re{3} nevertheless will easily be seen to be
valid non-perturbatively. 

\section{The Quantum Action Principle}
The Quantum Action Principle describes the 
response of a quantum field theory under a field transformation.
Thus it is a fundamental tool in the construction of field theories with
symmetry properties.
Let us consider an infinitesimal continuous transformation of the fields 
$\Phi$ of the theory 
\beq
\d\Phi_i(x)=\eps(x)\,P_i[\Phi(x)]\,,
\eeq
where the $P_i[\Phi]$ are (anticommuting)
polynomials in the fields, corresponding in the case of gauge theories to
BRS transformations, and $\eps$ is an
anticommuting parameter.
The path integral representation of the Euclidean generating functional is
\beq
Z[J,\eta]=\int {\mathcal D}\Phi \; e^{-S[\Phi,\eta]+J_i\Phi_i}\,,
\eeq
where (perturbatively)
\beq\nome{bare}
S[\Phi,\eta]=S_{\mbox{\footnotesize{classical}}}[\Phi]
+S_{\mbox{\footnotesize{counterterms}}}[\Phi]
-\eta_i P_i[\Phi]
\eeq
and a regulator of ultraviolet (UV) divergences is assumed.
By performing the field transformation with a constant $\eps$
we get the identity
\beq\nome{QAP1}
\int d^4 x \; J_i(x)\frac{\d Z[J,\eta]}{\d\eta_i(x)}=\int {\mathcal D}\Phi 
\; \D[\Phi,\eta] \, e^{-S[\Phi,\eta]+J_i\Phi_i}\,,
\eeq
where
\beq\nome{Delta}
\D[\Phi,\eta]=\int d^4 x \; \frac{\d^2 S(\Phi,\eta)}{\d\Phi_i(x)\d\eta_i(x)}
-\int d^4 x \; \frac{\d S[\Phi,\eta]}{\d\Phi_i(x)}
\frac{\d S[\Phi,\eta]}{\d\eta_i(x)}
\equiv {\mathcal B}_\Phi \cdot S\,.
\eeq
Then the response of the system is given by the insertion of a local
operator of dimension $4$ minus the dimension of the field $\Phi_i$
plus the dimension of the corresponding variation $P_i[\Phi]$.
When removing the regulator (UV limit) the l.h.s. is finite, at least in 
perturbation theory, and this ensures that also the insertion of
the operator $\Delta$ is finite in the UV limit.
In terms of the generating functional of 1PI Green functions, eq.~\re{QAP1}
reads
\beq\nome{QAP2}
\int d^4 x \; \frac{\d\G[\Phi,\eta]}{\d\Phi_i(x)}
\frac{\d\G[\Phi,\eta]}{\d\eta_i(x)}=[\D\G]\,,
\eeq
where $\G[\Phi,\eta]=-\log Z[J,\eta]+J_i\Phi_i$ and the r.h.s. stands
for the 1PI functions with the insertion of $\D$.

In general one is interested in solving the equation $\D=0$.
Notice that in perturbation theory $[\D\G]=\D+{\mathcal O}(\hbar)$, and
from this it follows that the insertion of $\D$ is local at the first order in
which $\D$ itself is non-vanishing. Since there are only a finite number
of local operators of the correct dimension, it follows
that the equation $\D=0$ is, order by order, a finite number of conditions, 
which can eventually be satisfied by fine-tuning \cite{ft} the 
parameters in the action \re{bare}. The relations \re{QAP1} and \re{QAP2},
together with locality of $\D$ as just described, is known as the QAP.

We want to generalize these concepts to an effective theory,
obtained from a more fundamental one after integrating the high 
energy degrees of freedom.
It is clear that such a procedure of integration will generate effective
non-local\footnote{by ``non-local'' we mean here a series of local
interactions of arbitrarily high numbers of derivatives. 
These interactions are sometimes referred as quasi-local.} 
interactions and also the field transformations will become non-local.
Thus it is not evident how the QAP can be obtained at the effective level.

\section{RG flow equations}
The main idea of Wilson RG \cite{W,P,G,B} 
is to consider an interacting field theory 
as an effective theory, that is to regard the high frequency modes of the 
fields of the theory as generating 
effective couplings for the low energy modes. 
In this picture one introduces an UV cutoff $\L_0$, which is a just a
tool to define the Green functions of the theory. 
For simplicity we will consider local interactions.
Then a scale $\L$ is introduced and the
frequencies between $\L$ and $\L_0$ are viewed as generating
interactions for the
frequencies lower than $\L$. The physical parameters are fixed at a 
physical scale $\L_R\leq\L$. In particular we will choose $\L_R=0$.

We will denote with $\Phi_a=\{\phi,\psi,\bpsi\}$ the fields of the 
theory (the $\phi$'s are commuting fields while the $\bpsi$, $\psi$ are
anticommuting, fermions or ghosts) and $J_a=\{j,\bchi,-\chi\}$ the 
corresponding sources. 
As in the previous section, the sources $\eta$ are coupled to the 
composite operators defining the symmetry transformations of the 
fields. Summations over internal indices are understood.
The free propagators are collected in the matrix
\beq
D^{-1}_{ab}=
\left(
   \begin{array}{ccc}
   D^{-1}_1 & 0 & 0  \\
   0 & 0 & -D^{-1}_2 \\
   0 & D^{-1}_2 & 0
   \end{array}
\right) \,,
\eeq
where $D_1$ and $D_2$ are the free propagators of the $\phi$ and $\psi$, 
$\bpsi$ respectively. The Euclidean generating functional is
\beq\nome{Z}
Z[J,\eta]=\int {\mathcal D}\Phi \, \exp{\left\{
-\half(\Phi, \,D^{-1}\Phi)_{0\L_0}
-\si[\Phi,\eta;\L_0]
+(J,\,\Phi)_{0\L_0}
\right\}}
\,,
\eeq
where the interaction action $\si$ contains local, renormalizable 
interactions and the sources for the variations of the fields, as in 
\re{bare}. We have introduced a cutoff scalar product
$$
(f,\,g)_{\L\L_0}\equiv
\int_p\, K^{-1}_{\L\L_0}(p)\,f_a(-p)\,g_a(p)
\,,
\;\;\;\;\;\;\;\;\;\;\;\;\;\;\;\;\;\;
\int_p \equiv \int \frac{d^4p}{(2\pi)^4}\,,
$$
where $K_{\L\L_0}(p)$ is a cutoff function which is one for 
$\L^2\le p^2\le\L_0^2$ and rapidly vanishing outside this interval.
In order to be rigorous, this function can be taken to be always different 
from zero and of class $C^\infty$ \cite{P}.
By integrating over the high energy modes one finds
\beq\nome{Z'}
Z[J,\eta]=N[J;\L,\L_0] \;\int {\mathcal D}\Phi \,
\exp{\biggl\{
-\half(\Phi, \,D^{-1}\Phi)_{0\L}
-\se[\Phi,\eta;\L,\L_0]
+(J,\,\Phi)_{0\L}
\biggr\}}
,
\eeq
where the coefficient $N$ is given by
$$
\log N[J;\L,\L_0] = \half (J, \,D J)_{0\L_0}
-\half (J, \,D J)_{0\L} \,.
$$
The functional $\se$ is the Wilsonian effective action and contains 
the effective interaction coming from the frequencies $p^2>\L^2$.
It is possible to show that this functional is equivalent to a generalization
of \re{Z}, in which the free propagators contain $\L$ as an infrared
cutoff \cite{BDM1,M}. 
We thus define the generating functional of the cutoff connected
Green function
\beq\nome{ZL}
e^{-W[J,\eta;\L,\L_0]}=\int {\mathcal D}\Phi \, 
\exp{\biggl\{
-\half (\Phi, \,D^{-1}\Phi)_{\L\L_0}
-\si[\Phi,\eta;\L_0]
+\int_p J\Phi
\biggr\}}
\,.
\eeq
Then we have
\beq\nome{WL}
\se[\Phi,\eta;\L,\L_0]-\half (\Phi,\, D^{-1}\Phi)_{\L\L_0} 
=W[K_{\L\L_0}^{-1} J',\eta;\L,\L_0] \,,
\eeq
where we introduced the useful source
\beq\nome{J'}
J'_a=D^{-1}_{ab} \Phi_b\,.
\eeq
Namely, apart for the tree level two point functions, the Wilsonian
effective action is the generating functional of the connected 
amputated cutoff Green function. As one expects, it is technically 
easier to study the Legendre transform of $W[J,\eta;\L,\L_0]$, which 
we call ``cutoff effective action'' and is a generalization
of the usual quantum effective action, since it contains the infrared
cutoff $\L$ in the free propagators \cite{We,BDM1,M}
\beq\nome{Leg}
\G[\Phi,\eta;\L,\L_0]=W[J,\eta;\L,\L_0]+\int_p J\Phi\,.
\eeq
In the limits $\L\to 0$ and \UV, 
one recovers the physical quantum effective action. Both these limits
can be taken in perturbation theory \cite{P,B,BDM1,MR,BT,BDM2}.
In particular the dependence on
the ultraviolet cutoff $\L_0$ will be sometimes understood.

We now come to the discussion of the $\L$-dependence of the Wilsonian and 
cutoff effective actions.
By derivating the corresponding definitions with respect to $\L$
one finds the following flow equations in $\L$, the Wilson or ``exact'' 
renormalization group equations. 
\beq\nome{erg1}
\LdL \se[\Phi,\eta;\L,\L_0]=- e^{\set} \left[ \half \LdL \left( 
\K \frac{\d}{\d\Phi},\,\K D\frac{\d}{\d\Phi} \right)_{\L\L_0}
\right]
e^{-\set} \,,
\eeq
\beq\nome{erg2}
\LdL \Pi[\Phi,\eta;\L,\L_0]= -\frac{1}{2} 
\int_q [\LdL K_{\L\L_0}^{-1}(q)] (-1)^{\d_a} D^{-1}_{ab}(q)
\left(\frac{\d^2\G}{\d\Phi_a(q)\d\Phi_b(-q)}\right)^{-1} 
\,,
\eeq
where $\d_a=1$ if $\Phi_a$ is a fermionic field and $0$ otherwise and
\beq\nome{pigreco}
\Pi=\G-\half (\Phi,\,D^{-1}\Phi)_{\L\L_0}+
\half (\Phi,\,D^{-1}\Phi)_{0\L_0}
\eeq 
is the cutoff effective action in which the infrared cutoff in the free 
propagators has been removed. 
In eq.~\re{erg2} the inverse of the second derivative of $\G$ is the matrix 
inverse taken in the space of indices $a,b$.
({\it N.B.} Functional derivatives are defined with respect to the measure
thus $\frac{\delta}{\delta J(q)} \int_p J\Phi =\Phi(q)$,  {\it etc.})

We see an essential feature: the flow equation for 
$\se$ contains in the r.h.s. terms of the same loop order as the l.h.s..
Thus in order to perform any perturbative study 
a filtration \cite{BBBCD}
(\ie the introduction of a field-counting operator) 
in the space of vertices is required and the analysis at any loop order
must be done by starting from the vertices with lower number of external 
fields.
However, these terms are 1-particle-reducible, so they disappear in the
flow equation for the cutoff effective action, thus rendering the latter
preferable in perturbation theory.

In order to integrate the RG equations \re{erg1}-\re{erg2} one has 
to supply the boundary conditions. 
For this reason it is useful to split the cutoff effective action into 
two parts. 
One performs a Taylor expansion of the cutoff vertices around vanishing 
momenta. If there are massless fields, the expansion must be done around 
non-vanishing subtraction points.
This expansion will have coefficients of decreasing dimension.
These coefficients are the couplings of the theory.
The ``relevant'' part is obtained by  keeping the terms with coefficients
having non-negative dimension (relevant couplings). 
All the remaining part is called ``irrelevant''. 
For instance in the scalar case one gets for the relevant part of the 
cutoff effective action 
$$
\Pirel[\phi;\L]=\frac 1 2 \int d^4x \; 
\phi(x)\bigl[ \s_1(\L)+\s_2(\L) \p^2 \bigr]
\phi(x)+\frac {\s_3(\L)}{4!} \int d^4x \;
\phi^4(x) \,.
$$
Since we expect the theory to be renormalizable, for $\L\sim\L_0$
the dimension of the irrelevant couplings should be given only by 
powers of $\L_0$. Thus the simplest boundary condition for the 
irrelevant part of the cutoff effective action is
\beq\nome{irr}
\Girr[\Phi,\eta;\L=\L_0]=0\,.
\eeq
Clearly, by using the boundary condition \re{irr}, there is no hope that
the cutoff effective action $\G[\Phi,\eta;\L,\L_0]$ will satisfy the QAP:
Since the cutoff $\L_0$ breaks the gauge symmetry, non-local symmetry 
breaking terms, proportional to inverse powers of $\L_0$,
will be generated in perturbation theory by the loop corrections.
Then a tremendous fine-tuning of the irrelevant vertices in
$\Girr[\Phi,\eta;\L=\L_0]$ is needed in order to cancel these non-invariant 
contributions.

Therefore with the boundary condition \re{irr} the cutoff effective 
action will fulfil the QAP only in the limit \UV.
The result of the RG method is a non-local effective action
$\G[\Phi,\eta;\L,\L_0\to\infty]$. In this sense, the introduction
of the UV cutoff $\L_0$ and the ``unphysical'' condition \re{irr}
on the effective action at the scale $\L_0$ are 
only technical tools to define the theory, while the goal of the 
procedure is the ``physical'' perturbative effective action at any 
scale $\L$.

For the relevant part it is useful to put the boundary conditions in 
the infrared, when most of the degrees of freedom have been integrated out
(in particular at the point $\L=0$, where the cutoff effective action becomes 
the physical one, so that the relevant couplings are related to 
measurable quantities). In the usual field theory language this means
giving the physical renormalization conditions. In the language of the
Wilson RG, this means that the flow in the infrared is controlled
by the relevant couplings. This is a highly non-trivial step of the
procedure, since at this point one really defines 
the theory, with its symmetries and physical couplings and masses.

We are now able to study the symmetry properties of the Wilsonian effective
action and cutoff effective action. Our aim is to implement the relation
\beq\nome{SJ}
{\mathcal S}_J \cdot Z[J,\eta]\equiv\int d^4 x\left[ 
j(x)\frac{\d}{\d\eta_1(x)}+\chi(x)\frac{\d}{\d\eta_2(x)}
-\bchi(x)\frac{\d}{\d\eta_3(x)}\right]
Z[J,\eta]=0\,,
\eeq
known as Slavnov-Taylor identities. This will be done in the next section.

\section{Effective Ward identities and locality}
We perform the following cutoff change of variables in the generating 
functional \re{Z'} \cite{B,BDM3}
$$
\d\Phi_a(p)=-\eps K_{0\L}(p) \frac{\d\se}{\d\eta_a(-p)}\,,
$$
where $\eps$ is a Grassmann parameter.
We get the identity
\beeq
&&{\mathcal S}_J \cdot Z[J,\eta]=N[J;\L,\L_0]\nonumber
\\
&&\quad\quad\times\nonumber
\int{\mathcal D}\Phi\;\De \; 
\exp{\biggl\{
-\half (\Phi, \,D^{-1}\Phi)_{0\L}
-\se[\Phi,\eta;\L,\L_0]
+(J,\,\Phi)_{0\L} 
\biggr\}}
\,,
\eeeq
where the operator giving the Ward identity violation at the 
effective level is
\beq\nome{Deltaeff}
\De=-{\mathcal S}_{J'} \cdot\se 
+\int d^4 p \; K_{0\L}(p)\;{\mathcal B}_\Phi \cdot\se\,,
\eeq
where ${\mathcal S}$, $J'$ and ${\mathcal B}$ have been defined in \re{SJ}, 
\re{J'} and \re{Delta}, respectively. 
This formula is completely analogous to \re{Delta}.
We see that the first term gives the usual Ward identities, while
the second is coming from the low momentum modes which still have to
be integrated out.
In order to get information about $\De$, in the following we will study 
in detail the properties of the flow of this operator. 
$\De$ satisfies a linear evolution equation 
(found by explicit derivation) \cite{B,BDM3}
\beeq\nome{flowDe}
\LdL\De &=& 
e^{2\set} \left[ \half \LdL \left(
K_{0\L} \frac{\d}{\d\Phi},\, K_{0\L} D
\frac{\d\De}{\d\Phi} \right)_{0\L}
\right] e^{-2\set} \,,
\nonumber \\ &=& \int_p \, [\LdL K_{0\L}(p)] \bigl\{
L_1+\hbar L_2
\bigr\} \De\,,
\eeeq
where the linear operators $L_1$ and $L_2$ are given by
\beeq\nonumber
&&
L_1
=D_1(p) \frac{\d\se}{\d\phi(-p)}\frac{\d}{\d\phi(p)}
-D_2(p) \frac{\d\se}{\d\psi(-p)}\frac{\d}{\d\bpsi(p)}
+D_2(p) \frac{\d\se}{\d\bpsi(-p)}\frac{\d}{\d\psi(p)}\,,\\
&& L_2=
-\half D_1(p) \frac{\d^2}{\d\phi(-p)\d\phi(p)}
-D_2(p) \frac{\d^2}{\d\psi(-p)\d\bpsi(p)}
\,.\nome{L}
\eeeq
In eq.~\re{flowDe} we restored the powers of $\hbar$ in order to show
how in the r.h.s. of the flow equation for $\De$ there are terms at the
same loop order of the l.h.s..  

Since $\De$ satisfies a linear equation, the gauge symmetry condition 
$\De=0$ is verified for any $\L$ if we can set to zero the boundary 
conditions of \re{flowDe}. The main point is to fix to zero the ones 
for the relevant part $\Der$ of $\De$ for some value $\L_R$ of the IR 
cutoff. 
Normally $\Der(\L_R)=0$ is a set of constraints which overdeterminates 
the couplings in $\se(\L_R)$. The number of independent constraints can be 
reduced by exploiting the so-called consistency conditions,
which are a set of algebraic identities coming from the anticommutativity 
of the differential operator $\frac{\d}{\d\eta}\frac{\d}{\d\Phi}$
\cite{B}. 
However, for this it is crucial the way in which the relevant parts are 
defined. If we keep the IR cutoff $\L_R\ne 0$, 
we can extract $\Der(\L_R)$ by expanding 
the vertices of $\De(\L_R)$ around vanishing momenta, even though we 
are considering massless particles. The result is that the consistency 
conditions constrain some couplings in $\Der(\L_R)$, so that the set 
$\Der(\L_R)=0$ can be fulfilled in some cases by tuning 
the parameters in $\ser$. 
See ref.~\cite{B} for such an analysis in the pure gauge SU(2) model. 
If we are interested in fixing the boundary conditions at the physical 
point $\L_R=0$ in a theory with one or more massless particles, we 
have to introduce non-vanishing subtraction points in order to define 
$\Der(\L_R=0)$. This fact could spoil the power of the consistency 
conditions since they now involve also irrelevant vertices of $\De(0)$ 
evaluated at the subtraction points \cite{BDM3}. 
Thus it seems that a case-by-case analysis based on a filtration of $\De$ 
is required in order to prove the locality of $\De$ so as to restore 
the usefulness of the consistency conditions.
This was done in ref.~\cite{BDM3}.

However these nasty irrelevant contributions are of the reducible type 
\cite{BDM3} (see the form of $L_1$ in \re{L}), and we 
expect they will disappear when taking the Legendre transform,
similarly to what happened in passing from the Wilsonian effective action 
to the cutoff effective action.

{F}rom \re{WL} and \re{Deltaeff}, the expression of $\De$ in terms of 
$W[J,\eta;\L,\L_0]$ is 
\beq
\De=-{\mathcal S}_{K_{0\L_0}J} \cdot W[J,\eta;\L,\L_0]
+\int d^4 p \;
\frac{K_{0\L}(p)}{K_{\L\L_0}(p)}  
\;{\mathcal B}_{J''} \cdot W[J,\eta;\L,\L_0]\,,
\eeq
where $J''=DJ$. By performing the Legendre transform \re{Leg}
one gets the cutoff Ward identities \cite{E} ($\DG$ is $\De$ expressed
in terms of $\Phi$):
\beeq
&&\DG=-\int d^4 p \; K_{0\L_0}(p)
\frac{\d\G}{\d\Phi_a(-p)}\frac{\d\G}{\d\eta_a(p)}
+\int_p \frac{K_{0\L}(p)}{K_{\L\L_0}(p)}  (-1)^{\d_a}
D^{-1}_{ab}(p) \\&&\quad\times
\left[\Phi_a(p)\frac{\d\G}{\d\eta_b(p)}
+\int_q \left(\frac{\d^2 \G}{\d\Phi_a(p) \d\Phi_c(q)}\right)^{-1}
\frac{\d^2 \G}{\d\Phi_c(-q)\d\eta_b(-p)}
\right]\,.\nonumber
\eeeq
The expression of $\DG$ is simpler in terms of the functional $\Pi$,
defined in \re{pigreco}. One finds
\beeq\nome{cutoffST}
&&\D_\G=-\int d^4 p \; K_{0\L_0}(p) 
\frac{\d\Pi}{\d\Phi_a(-p)}\frac{\d\Pi}{\d\eta_a(p)}\\
&& \quad\quad
+\int_{p,q} \frac{K_{0\L}(p)}{K_{\L\L_0}(p)} (-1)^{\d_a}
D^{-1}_{ab}(p) 
\left(\frac{\d^2 \G}{\d\Phi_a(p) \d\Phi_c(q)}\right)^{-1}
\frac{\d^2 \Pi}{\d\Phi_c(-q)\d\eta_b(-p)}
\,.\nonumber
\eeeq
Again the flow equation for the cutoff ST is found by explicit 
derivation. We have
\beeq\nome{evST} 
&&\LdL \D_\G= \int_{p,q,r}
[\LdL K_{\L\L_0}^{-1}(p)] (-1)^{\d_d} D_{ab}^{(-1)}(p)
\\ && \quad\quad\quad \times\nonumber
\half
\left(\frac{\d^2 \G}{\d\Phi_a(p) \d\Phi_c(q)}\right)^{-1}
\frac{\d^2 \D_\G}{\d\Phi_c(-q) \d\Phi_d(r)}
\left(\frac{\d^2 \G}{\d\Phi_d(-r)\d\Phi_b(-p)}\right)^{-1}
\,.
\eeeq
We see how this equation has the desired 
property, namely the evolution of the vertices of $\DG$ depends on
vertices of $\DG$ itself at lower loop order with respect to the 
l.h.s..

We now come to the discussion of the locality of $\D_\G$ in perturbation 
theory, which is an immediate consequence of eq.~\re{evST}. 

If one assumes that $\DG^{(\ell')}$ is vanishing at any loop order 
$\ell'<\ell$, then at loop $\ell$ $\DG^{(\ell)}$ is $\L$-independent, 
since the flow equation \re{evST} becomes
$$
\LdL\DG^{(\ell)}=0\,.
$$
Then we choose to discuss the locality of $\DG^{(\ell)}$ (given in 
\re{cutoffST}) at the point $\L=\L_0$. At this point the functionals 
$\G$ and $\Pi$ are local, as follows from \re{irr}, and the free 
propagators vanish, so that also from 
$\left(\frac{\d^2 \G}{\d\Phi_i(p) \d\Phi_j(q)}\right)^{-1}$
only local terms are coming.
Thus the only possible source of non-local contributions is the cutoff 
function $K_{0\L_0}(p)$. It is then sufficient to take the UV limit
\UV, in which $K_{0\L_0}\to 1$, to get a local violation 
$\DG^{(\ell)}$.

Once the locality of $\DG^{(\ell)}$ is proven,
powerful cohomological methods \cite{B} can be 
used to translate the finite set of equations $\DG^{(\ell)}=0$ into a 
solvable set of fine-tuning conditions on the couplings of $\Pirel^{(\ell)}$.

\section{Final remarks}
Now consider non-perturbative approximations to the 
flow equations. For the reasons stated in the introduction, we 
do not accept approximations which destroy gauge invariance. 
This means that eqn.~\re{flowDe} or equivalently \re{evST} must be absolutely
respected. The problem is that even if $\se$ and thus $\De$ (or, $\G$
and thus $\DG$) are polynomial in the fields and/or polynomial in momenta
at the cutoff scale $\Lambda_0$, this
is not preserved by the flows \re{flowDe}  
[respec. \re{evST}].  
Thus any succesful approximation of the flow equations
would have to involve a non-polynomial 
non-local action -- ruling out in particular, any 
straightforward truncation.
The underlying difficulty is that the
effective BRS transformation(s) (\ie the dependence on $\eta$) is
also non-polynomial and non-local.
In principle, knowledge of this effective transformation 
would help determine an appropriate form of approximations
to the effective action, but of course the form of the effective BRS
transformation is also unkown. By differentiating \re{Deltaeff} with
respect to $\Phi$ and $\eta$ (and recalling that 
$\frac{\d}{\d\eta}\frac{\d}{\d\Phi}$ is Grassmann)
a set of effective Wess-Zumino consistency conditions may be derived
which constrain the dependence on $\eta$. However these are no easier to solve
directly than the flow equations. We conclude that the loss of the
locality of the QAP in the non-perturbative domain, causes severe
difficulties for all approximations that use method (a).

In spite of these conceptual difficulties in studying a truncation 
of the effective action, we should mention the
approach followed by the authors in ref.~\cite{E}. They accept the
unavoidable violation to the Ward identities caused by a given truncation of
the effective action. However, they use the flow equation for the 
(truncated) effective Ward identity operator $\De$ to study
numerically the deviation of such an approximated effective action
from the gauge invariant trajectory. In this way they obtain a check
of the consistency of the given truncations.

We are grateful for discussion to M. Asorey, C. Becchi, M. Bonini, 
U. Ellwanger, M. Hirsch, G. Marchesini and A. Weber.

\end{document}